\documentclass[utf8]{frontiersFPHY} 

\setcitestyle{square} 
\usepackage{url,hyperref,microtype}
\usepackage[onehalfspacing]{setspace}
\usepackage{multirow}
\usepackage{float} 

\usepackage{graphicx}

\def\keyFont{\fontsize{8}{11}\helveticabold }
\def\firstAuthorLast{C. Neub\"user {et~al.}} 
\def\Authors{C. Neub\"user\,$^{1,*}$,  T. Corradino\,$^{1,2}$, G-F. Dalla Betta\,$^{1,2}$, L. De Cilladi\,$^{3,4}$, and L.~Pancheri\,$^{1,2}$ on behalf of the ARCADIA collaboration}
\def\Address{$^{1}$TIFPA-INFN, 38123 Trento, Italy \\
$^{2}$University of Trento, Department of Industrial Engineering, 38123 Trento, Italy\\
$^{3}$INFN-Sezione di Torino, 10125 Torino, Italy \\
$^{4}$University of Torino, Department of Physics, 10125 Torino, Italy}
\def\corrAuthor{Coralie Neub\"user}

\def\corrEmail{coralie.neubueser@tifpa.infn.it}

\begin{document}
\onecolumn
\firstpage{1}

\title[Sensor design optimizations of fully depleted FD-MAPS]{Sensor design optimization of innovative low-power, large area FD-MAPS for HEP and applied science} 

\author[\firstAuthorLast ]{\Authors} 
\address{} 
\correspondance{} 

\extraAuth{}

\maketitle

\begin{abstract}
Fully Depleted Monolithic Active Pixels (FD-MAPS) represent a state-of-the-art detector technology and profit from a low material budget and cost for high energy physics experiments and other fields of research like medical imaging and astro-particle physics. Compared to the MAPS currently in use, fully depleted pixel sensors have the advantage of charge collection by drift, which enables a fast and uniform response overall the pixel matrix. 
The functionality of these devices has been shown in previous proof-of-concept productions. In this article we describe the optimization of the test pixel designs, that will be implemented in the first engineering run of the demonstrator chip of the ARCADIA project. These optimization procedures include radiation damage models, that have been employed in Technology Computer Aided Design simulations to predict the sensors behavior in different working environments.
 
\section{}

\tiny
 \keyFont{ \section{Keywords:} MAPS, radiation detectors, pixel detectors, CMOS, TCAD simulations} 
\end{abstract}


\section{Introduction}
Silicon pixel detectors generally consist of a matrix of sensing diodes connected to readout electronics. Currently, there are two major types of technology used to implement this class of devices. First, hybrid detectors, that have the electronics integrated readout circuit on a separate die from the sensing active silicon substrate. In these devices each pixel is connected via bump bonding to the readout circuitry~\cite{WERMES2009370}. Secondly, monolithic detectors, that have the electronics integrated in the same Si substrate according to three main technological realizations; in (semi-monolithic) Depleted Field Effect Transistors (DEPFET) pixels~\cite{KEMMER1987365,VELTHUIS2007685}, Silicon-On-Insulator (SOI) active pixels~\cite{NIEMIEC20051202} and CMOS monolithic active pixels~\cite{TURCHETTA2001677,TURCHETTA2003251}. 
The DEPFET pixel technology offers very low-noise characteristics due to the small capacitance of the collection node. It is used in the Belle II experiment~\cite{MARINAS201331} as well as light source facilities, and astronomy~\cite{LUTZ2017122}. The limitation of this technology lies in the necessary discharge of the collected electrons below the collection node, which requires a comparatively low occupancy and low radiation environments, and the necessity of additional external circuits for control and readout. 
The SOI sensors use a thin low-resistivity Si layer for the circuit implementation and high-resistivity bottom wafer. These two layers are separated by a buried oxide layer, which enables thin substrate thicknesses and small pitches with low capacitance~\cite{Wang2015SoI}. However, these devices suffer from high sensitivity to radiation mainly due to the accumulation of positive oxide charges in the buried oxide layer~\cite{HAGINO2020164435}. 

\begin{figure}[htbp]
\centering
	\begin{minipage}{.95\textwidth}
		\includegraphics[width=1\textwidth]{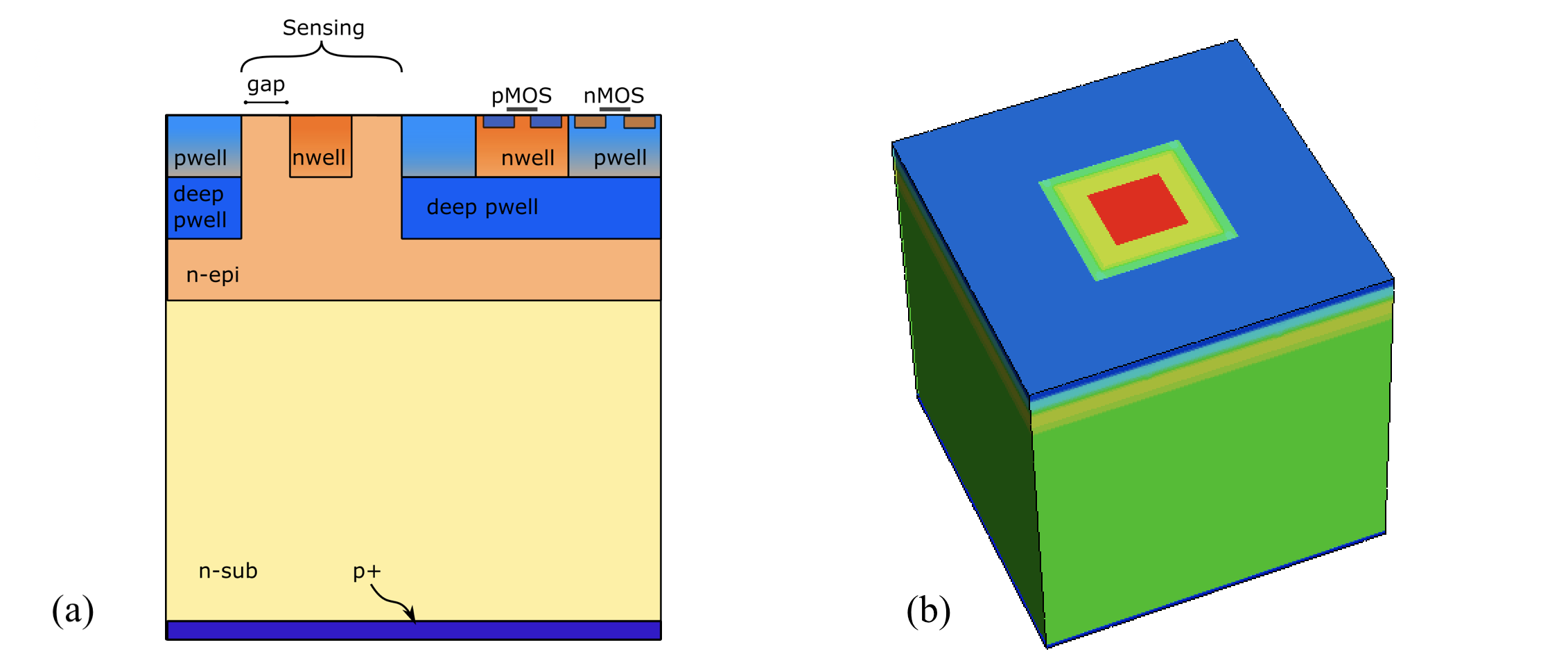}
	\end{minipage}
\caption{(a)~Schematic of the ARCADIA pixel. (b)~TCAD 3D single pixel simulation domain. The color scale refers to the doping concentration.}
\label{fig:scheme}
\end{figure}

Recent advancements in CMOS Monolithic Active Pixel Sensors (MAPS) have demonstrated their ability to operate in high radiation environments of up to multiple kGy's~\cite{SNOEYS201790}, which increased their appeal as sensors for high-energy physics detectors. The most recent example in such application is the new ALICE Inner Tracking System~\cite{MAGER2016434}, entirely instrumented with CMOS MAPS, that covers an area of about 10\,m$^{2}$.
However, the full potential of such devices has not yet been fully exploited, especially in respect of the size of the active area, power consumption, and timing capabilities. Ongoing developments concentrate on FD-MAPS, that profit from a charge collection by drift~\cite{PERIC2007876,Havr_nek_2015,GIUBILATO201391}. \\
The ARCADIA project is developing FD-MAPS with an innovative sensor design, which uses backside bias to improve the charge collection efficiency and timing over a wide range of operational and environmental conditions. The sensor design targets very low power consumption, in the order of 20\,mW cm$^{-2}$ at 100\,MHz cm$^{-2}$ particle flux, to enable air-cooled operation. Another key design parameter is the ability to further reduce the power regime of the sensor, down to 5\,mW cm$^{-2}$ or lower, for low hit rates like e.g. in space applications. The FD-MAPS architecture, initially embodied in a 512$\times$512 pixel matrix, should enable the scalability of the sensor up to matrix sizes of 2048$\times$2048 pixels. Maximizing the active area of the single sensor (10\,cm$^{2}$ or bigger) simplifies and reduces the costs of detector construction, and even enables applications where no support material over the entire sensor area can be tolerated (e.g. medical scanners). \\
The sensor design is based on a modified 110\,nm CMOS process and incorporates a low-doped n-type silicon active volume with a p$^{+}$ region at the bottom (Fig.~\ref{fig:scheme}). This structure will be obtained using two alternative approaches depending on the thickness of the active volume. For thick sensors, exceeding 100\,$\mu$m, the fabrication will be based on  high-resistivity n-type substrates, and the p+ region will be formed by back-side processing. For sensors thinner than 100\,$\mu$m, on the contrary, low-doped n-on-p+ epitaxial substrates will be used. The p-n junction sits on the bottom of the sensor, which results in the depletion region growing from the backside surface with increasing bias voltage. These FD-MAPS are thus operational at low front-side supply voltages while providing a fully depleted silicon bulk, which allows the electrode on the top to read out fast electron signals produced by drift. A more detailed description of the processing is provided in~\cite{pancheri2019110}. \\
The ARCADIA collaboration is currently working on a large-area prototype of 1.3$\times$1.3\,cm$^{2}$ active area consisting of $512\times512$ pixels with 25\,$\mu$m pitch, to be realized in a first engineering run with integrated digital electronics. Additional test structures of pixel matrices with pixel pitches ranging from 10 to 50\,$\mu$m and total thicknesses of 50 to 300\,$\mu$m will also be produced. 
In the following we report on the simulations, and the optimization procedures of the sensor designs that have been developed in preparation of the production run. 

This paper is organized as follows: Section 2 describes the pixel geometry and the simulation settings, that were used in the analysis of the device characteristics. Additionally, the radiation damage models used for the predictions after irradiation are introduced and the first observations discussed. The optimization targets, methods, and results are described in Section 3. The conclusions are given in Section 4.

\section{TCAD simulations}
The electrical characteristics of the pixels are analyzed using Technology Computer Aided Design (TCAD) simulations~\footnote{Synopsys Sentaurus Version P-2019.03-SP1}.
The TCAD simulation domains are realized in 3D, as visualized for a single pixel domain with 50\,$\mu$m thickness, in Fig.~\ref{fig:scheme}~(b). The collection electrode is biased at $+0.8$\,V and the reverse bias voltage is applied from the back side of the pixel. Three different substrate thicknesses of 50,100, and 300\,$\mu$m are simulated, using the same  doping profiles provided by the foundry. This basic simulation setup has been previously validated with a production series, under the SEED project~\cite{Pancheri2020}. 

All simulations, if not stated otherwise in the figures, are performed at 300\,K. \\
The simulations neglect the nwell regions used in the implementation of PMOS transistors, assuming that the electrical characteristic and the radiation detection is not significantly affected. \\
Due to the wide range of planned applications, these studies include a detailed analysis of the impact of different radiation levels on the pixel characteristics using radiation damage models for both the silicon bulk and the Si-SiO$_{2}$ interface. The adopted models, as well as the simulation results, are discussed in Sec.~\ref{subsec:rad}.\\

\subsection{Electrical characterization}
\label{subsec:Vop}
\begin{figure}[htbp]
\begin{center}
	\includegraphics[width=.9\textwidth]{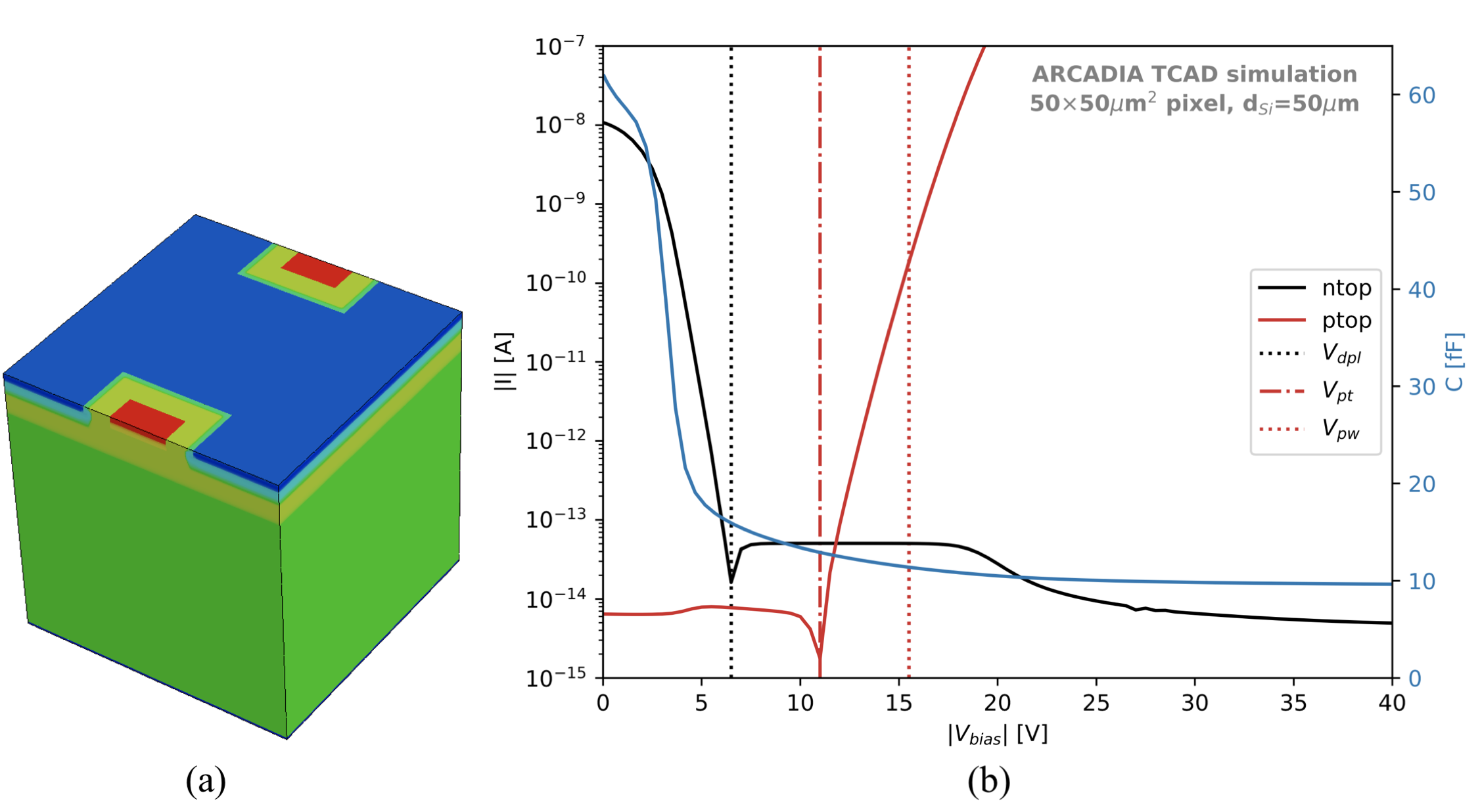}
\caption{(a)~Example of a 3D TCAD simulation domain of two half-pixels. The color scale refers to the doping concentration.(b)~Simulated IV and CV curves of the standard 50$\times$50\,$\mu$m$^2$ pixel  in $d_{Si}=50$\,$\mu$m thickness. The black curve shows the absolute current measured at the collection electrode, the red curve corresponds to the current measured at the pwell on the top side of the sensor. The blue curve shows the capacitance of the full structure over the applied bias voltage range. The dotted black, dashed red, and dotted red lines mark the voltages of depletion $V_{dpl}$, onset of the punch-through current $V_{pt}$, and reached maximum power consumption $V_{pw}$, respectively.}
\label{fig:example}
\end{center}
\end{figure}

An example of the capacitance-voltage (CV) and current-voltage (IV) curves of a 50$\times$50\,$\mu$m$^2$ pixel is shown in Fig.~\ref{fig:example}~(b). The CV characteristics are, if not stated differently, simulated in AC with a frequency of 10\,kHz. 
The black curve in Fig.~\ref{fig:example}~(b) represents the current measured at the collection diode and has been obtained from a two half-pixel domain, shown in Fig.~\ref{fig:example}~(a), with a 10\,mV voltage difference applied at the top electrodes. In this way, a non-negligible current is flowing between the pixels if the substrate is not fully depleted. The red and blue curves correspond to current measured at the top pwell, and capacitance measured at the collection electrode in single pixel simulations, respectively. The dotted and dashed lines highlight the depletion voltage and onset of the punch-through currents, respectively. Both IV curves feature the point of sign change from negative to positive current values in form of dips in their absolute spectrum. These are defined as the voltage needed for full depletion $V_{dpl}$ and onset of the punch-through currents $V_{pt}$, and correspond to the moment in which the single pixels become isolated from each other through the depletion region, and the moment in which the exponential increase of hole current between the pwell implantations and the back side p$^{+}$ contact starts.  
The sensors can be biased at a voltage higher than $V_{pt}$, as long as the power consumption is not too large. Therefore, it is useful to extract the maximum allowed bias voltage at which a power consumption of 0.1\,mW/cm$^2$ is reached. This voltage is marked as dotted red line, and will be referred to as $V_{pw}$ in the following. \\
As it is visible in Fig.~\ref{fig:example}~(b), comparing the black dotted line and blue curve, the depletion voltage of these devices does not necessarily correspond to the voltage at which the minimum capacitance is reached. Due to the low doped epitaxial layer, full depletion of the substrate including the epi-layer, is only achieved at much higher voltages. 

In the following analysis, capacitances and leakage currents are extracted at $V_{pt}$, which we have chosen as an optimal and safe operating point.

\subsection{Prediction of radiation tolerance}
\label{subsec:rad}

The ARCADIA sensors are designed for a wide range of applications, which include HEP collider experiments. Future collider experiments like the FCC, require good timing performance and (partially) high radiation tolerance, especially for the central tracking and vertex detectors~\cite{Abada:2019ab,Abada:2019aa}. \\ 
Ionizing radiation not only damages the surface of the sensor, which produces oxide charges at the interface between the Si and SiO$_{2}$ surface, but creates cluster and point-like defects in the silicon crystal lattice. These defects result in new energy levels within the band-gap and can function as traps which get electrically active when occupied, thus changing the electrical properties of the sensors, such as the depletion voltage (effective doping concentration) and the leakage (or dark) current. \\
For a most realistic prediction of the impact of these defects, we have modeled both damage types, following the so called \textit{new Perugia model}~\cite{Morozzi:2020yxm,Passeri:2705944}. This three-trap model introduces two acceptor and one  donor state in the bandgap for the description of non-ionising radiation damage in the silicon bulk. In addition, two trap states at the Si-SiO$_{2}$ interface are introduced to describe the effects of ionising energy losses, together with fixed oxide charges which build up due to trapped charges in the SiO$_{2}$ layer. \\
While this model has been tuned and validated on p-type silicon, it has shown good performance for sensors produced by different vendors~\cite{Morozzi:2020yxm}. These results make us confident that the model has a good qualitative predictive power also for the ARCADIA sensors. However, a variation in absolute values of leakage current and capacitance can be expected due to varying impurity concentrations in the substrate.\\

\subsubsection{Surface damage}
\label{subsec:surfDamage}
\begin{figure}[htbp]
\begin{center}
	\includegraphics[width=.65\textwidth]{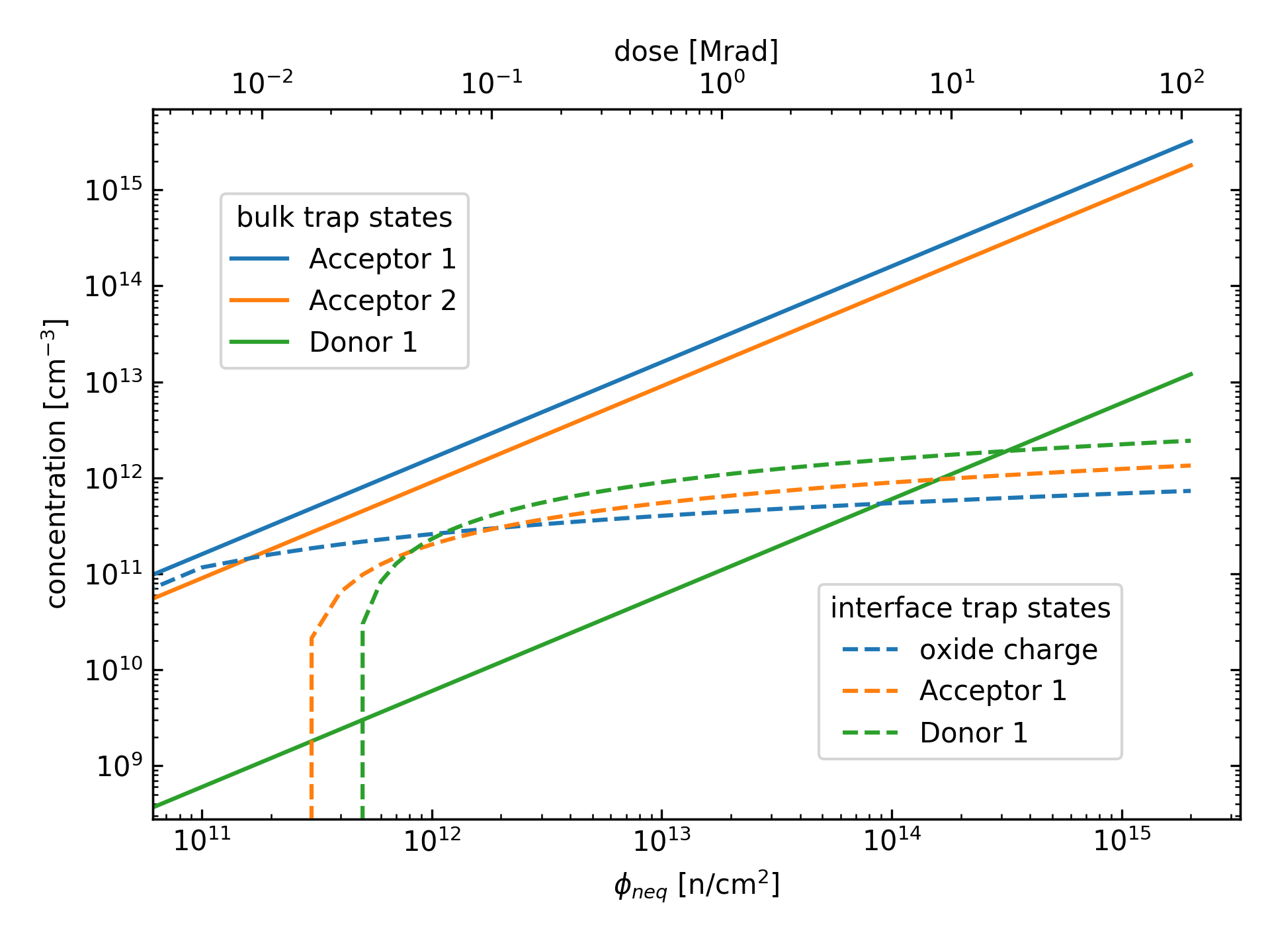}
\caption{Trap concentrations as a function of the neutron equivalent fluence (bottom axis) and total ionizing dose (top axis)~\cite{Passeri:2705944}. The solid and dashed lines correspond to the trap states in the silicon bulk and at the Si-SiO$_{2}$ interface, respectively. The trap properties are listed in Table~\ref{tab:traps}.}
\label{fig:concentrations}
\end{center}
\end{figure}

To estimate the impact of surface damage on the sensor properties, we have employed a SiO$_{2}$ layer that features positive oxide charges. The concentration of these charges increases with the modeled total ionizing dose (TID), along with defect levels in the band gap. We are following the model introduced and summarized in~\cite{Passeri:2705944}.
The trap and oxide charge concentrations are plotted as a function of the 1\,MeV neutron equivalent fluence and the dose in Fig.~\ref{fig:concentrations}. Here, the fluence dose relation is used for 24\,GeV protons of the CERN Proton Synchrotron with a proton hardness factor of 0.6 and a damage factor of $D/\phi=3.3\times10^{-8}\,$rad$/$(proton/cm$^{2}$)~\cite{Mandic2004}.
Because most potential applications of the ARCADIA technology do not expect doses greater than $10$\,krad, this value has been chosen as a reference for capacitance and leakage current estimates.  

\begin{figure}[htbp]
\begin{center}
\centering
	\includegraphics[width=\textwidth]{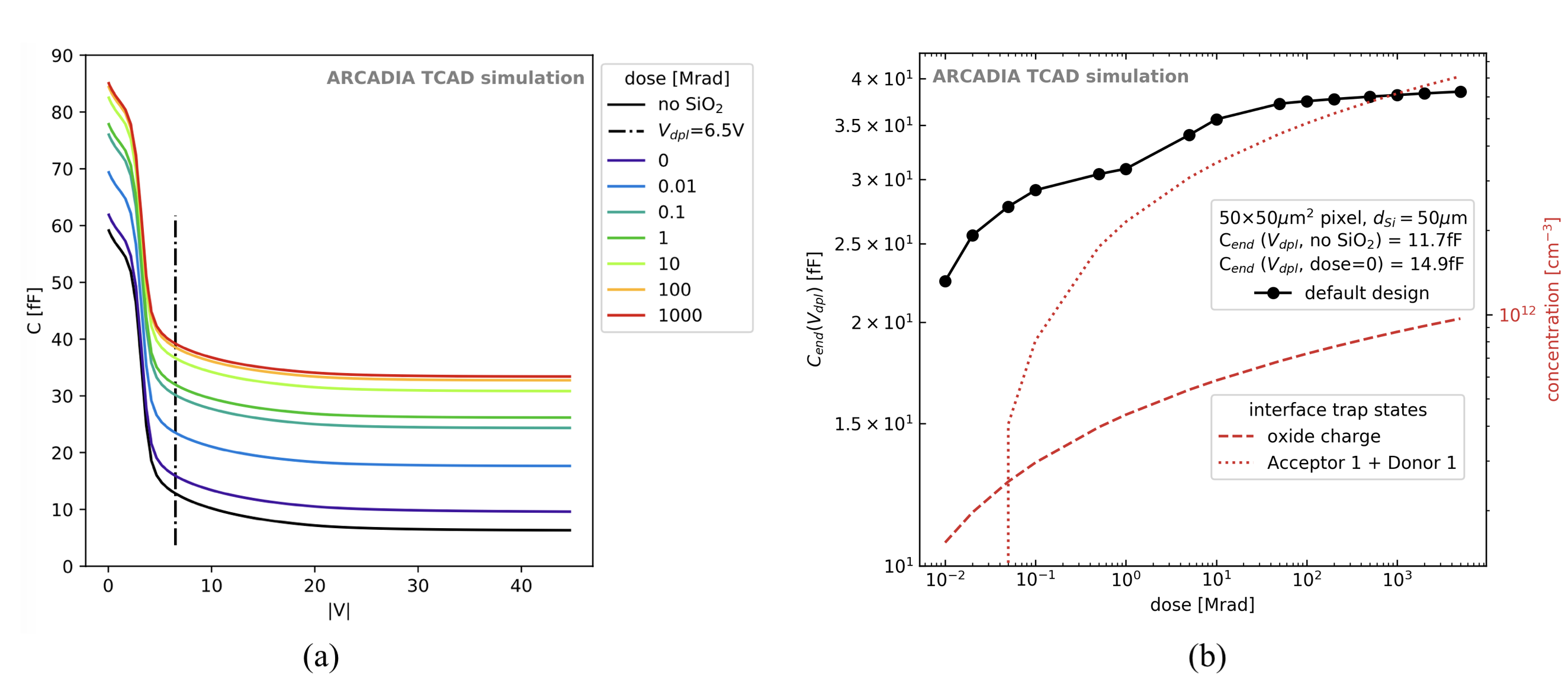}
\caption{(a)~Capacitance-voltage characteristics of a 50$\times$50\,$\mu$m$^2$ pixel in $d_{Si}=50$\,$\mu$m thickness, for different total ionising doses. The dashed line corresponds to the depletion voltage of $V_{dpl}=6.5$\,V. (b)~The capacitance at the depletion voltage $V_{dpl}$ of the 50$\times$50\,$\mu$m$^2$ pixel as a function of the total ionising dose. The red dashed and dotted lines correspond to the concentration of positive oxide charges and the sum of the two trap concentrations at the Si-SiO$_{2}$ interface, respectively. }
\label{fig:surface}
\end{center}
\end{figure}

The impact of the SiO$_{2}$ and of the induced positive oxide charges is visible in the CV curves, as presented in Fig.~\ref{fig:surface}~(a) of the example of a 50$\times$50\,$\mu$m$^2$ pixel. An increase of the capacitance is observed over the full voltage range (from 0 to -50V). This effect is present for all the pixel pitches. The CV curve without the SiO$_{2}$ layer is represented by the black line and shows the smallest values. When the SiO$_{2}$ layer is included, with a minimal concentration of positive oxide charge and traps, an increase of about 3\,fF at the depletion voltage (-6.5\,V) is observed. With increasing dose, of up to 5\,Grad (50\,MGy), the capacitance more than doubles. The first increase of the capacitance at the depletion voltage $C_{end}(V_{dpl})$ for low doses ($<$1\,Mrad), a second increase after 1\,Mrad, and a saturation above 100\,Mrad is shown in Fig.~\ref{fig:surface}~(b). The two steps can be directly correlated to the positive oxide charge, and trap concentrations at the Si-SiO$_{2}$ interface, as shown in red dashed and dotted lines, respectively.

\begin{figure}[htbp]
\begin{center}
	\includegraphics[width=.8\textwidth]{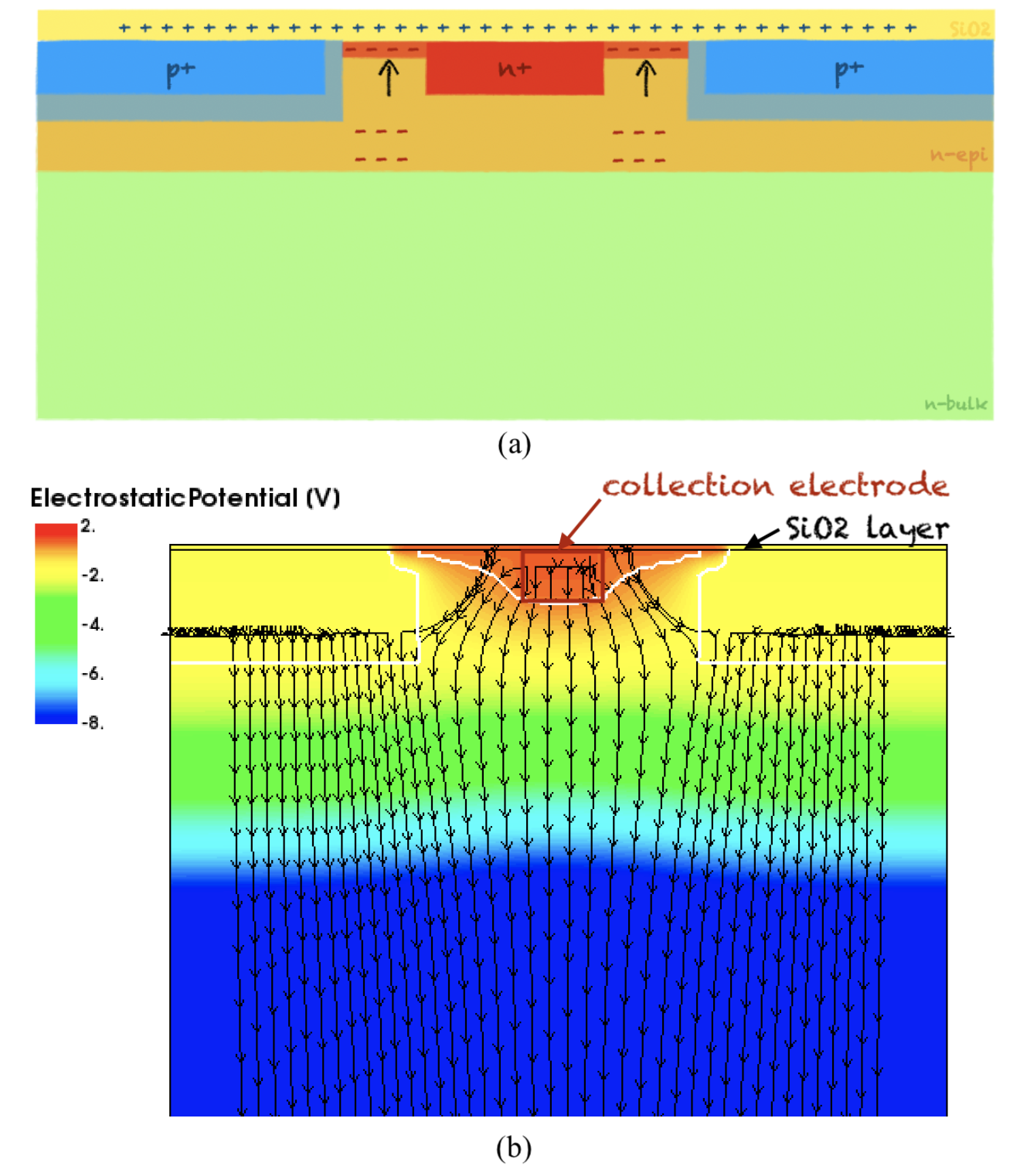} 
\caption{(a)~Schematic of impact of positive oxide charges at the interface between silicon and SiO$_{2}$. (b)~Electrostatic potential across the center of a 25$\times$25\,$\mu$m$^2$ pixel including a SiO$_{2}$ layer and positive oxide charges corresponding to 50\,Mrad total ionising dose. The black arrows show the electric field lines, and the white lines mark the depletion region.}
\label{fig:oxide}
\end{center}
\end{figure}

The main impact of the increased trap concentration is a higher surface recombination velocity, while the positive oxide charges attract free electrons from the n-type epi layer and create an effective extension of the collection electrode. This effect is schematically visualized in the top plot of Fig.~\ref{fig:oxide}~(a), and its impact on the electrostatic potential and the electric field lines is shown in Fig.~\ref{fig:oxide}~(b). \\
The strength of this effect depends on the sizes of the collection electrode and of the implant-free gap size between the diode and the pwell of the pixel. This allows for an optimization of the geometry and will be further discussed in Sec.~\ref{subsec:minC}.\\

\subsubsection{Bulk damage}
\label{subsec:bulkDamage}
To simulate the pixels' electrical characteristics after irradiation with hadrons, we have employed a three trap model, similarly to~\cite{FOLKESTAD201794, Moscatelli2016}. The main properties like the activation energy, and the capture/emission cross-sections for electrons and holes of the defect levels are summarized in Table~\ref{tab:traps}.

\begin{table}[htp]
\begin{center}
\begin{tabular}{c|c|c|c|c|}
		&		& energy level	 & $\sigma_{e}$ 		& $\sigma_{h}$ \\
material	& name 	& [eV]		 & [cm$^{2}$]			& [cm$^{2}$] \\
\hline
\multirow{2}*{SiO$_{2}$} & Acceptor 1		&	$E_{C}-0.56$	& $1\times10^{-16}$		& $1\times10^{-15}$\\
	 	& Donor 1			&	$E_{V}+0.6$	& $1\times10^{-15}$		& $1\times10^{-15}$\\
\hline
		& Acceptor 1		& $E_{C}-0.42$	 	& $1\times10^{-15}$		& $1\times10^{-14}$ \\
Si bulk	& Acceptor 2		& $E_{C}-0.46$ 	& $7\times10^{-14}$		& $7\times10^{-13}$ \\
		& Donor 1 		& $E_{C}-0.23$	 	& $2.3\times10^{-14}$	& $2.3\times10^{-15}$ \\
\hline
\end{tabular}
\end{center}
\caption{Properties for the two surface and three bulk traps~\cite{Passeri:2705944}; listed as in which material they are introduced, how they are named in Fig.~\ref{fig:concentrations}, which energy level in the Si bandgap they occupy, and their electron $\sigma_{e}$ and hole $\sigma_{h}$ capture cross-sections, respectively. }
\label{tab:traps}
\end{table}%

\begin{figure}[htbp]
\begin{center}
	\includegraphics[width=.59\textwidth]{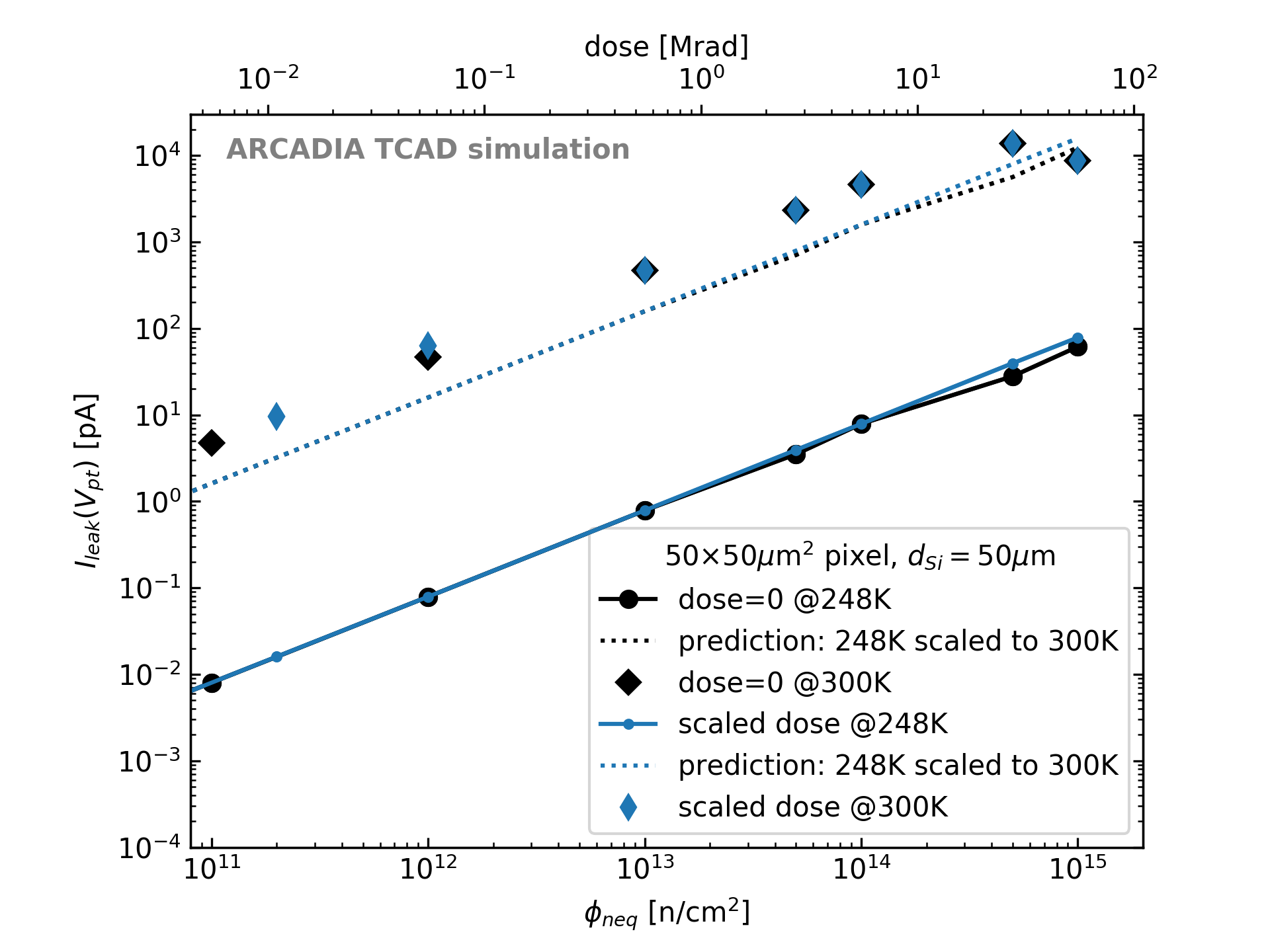}
\caption{Leakage current of a 50$\times$50\,$\mu$m$^2$ pixel in $d_{Si}=50$\,$\mu$m thickness as a function of the fluence. The black markers and lines correspond to a fixed dose of zero, while the blue markers/lines show the results for increasing concentrations of traps and oxide charges at the Si-SiO$_{2}$ interface. The circular markers correspond to simulations run at 248\,K and the diamonds show the results at 300\,K. }
\label{fig:Ileak_fluence}
\end{center}
\end{figure}

The concentration of these traps increases linearly with the fluence, as shown in Fig.~\ref{fig:concentrations}, and this results in a linear increase of the leakage current. This effect is shown in Fig.~\ref{fig:Ileak_fluence}, together with the comparison of a simple scaling of the current from low ($T_{1}=248$\,K) to higher ($T_{2}=300$\,K) temperatures, following~\cite{Sze}:
\begin{equation}
\label{eq:Ileak}
\frac{I(T_{2})}{I(T_{1})}=\left(\frac{T_{2}}{T_{1}}\right)^{2}\exp\left(-\frac{E_{g,eff}}{2\cdot k_{B}}\left[\frac{1}{T_{2}}-\frac{1}{T_{1}}\right]\right)
\end{equation}
with $E_{g,eff}=1.214$\,eV. Fig.~\ref{fig:Ileak_fluence} shows the measured leakage current at $V_{pt}$ as a function of the fluence $\phi_{neq}$ for 50$\times$50\,$\mu$m$^2$ pixels in 50\,$\mu$m thickness. The first observation is, that the induced surface damage has no significant impact on the leakage current, which is dominated by the volume generation currents induced by the bulk defects.\\ 
The second observation is made using Equation~\ref{eq:Ileak} to scale the values measured at 248\,K up to room temperature (shown in dotted lines). Since the model parameters are fitted to measurements at 248\,K, it has to be mentioned that we observe a difference between the simulation results at 300\,K and the scaling from 248\,K, of a factor three. 

\begin{figure}[htbp]
\begin{center}
	\includegraphics[width=.59\textwidth]{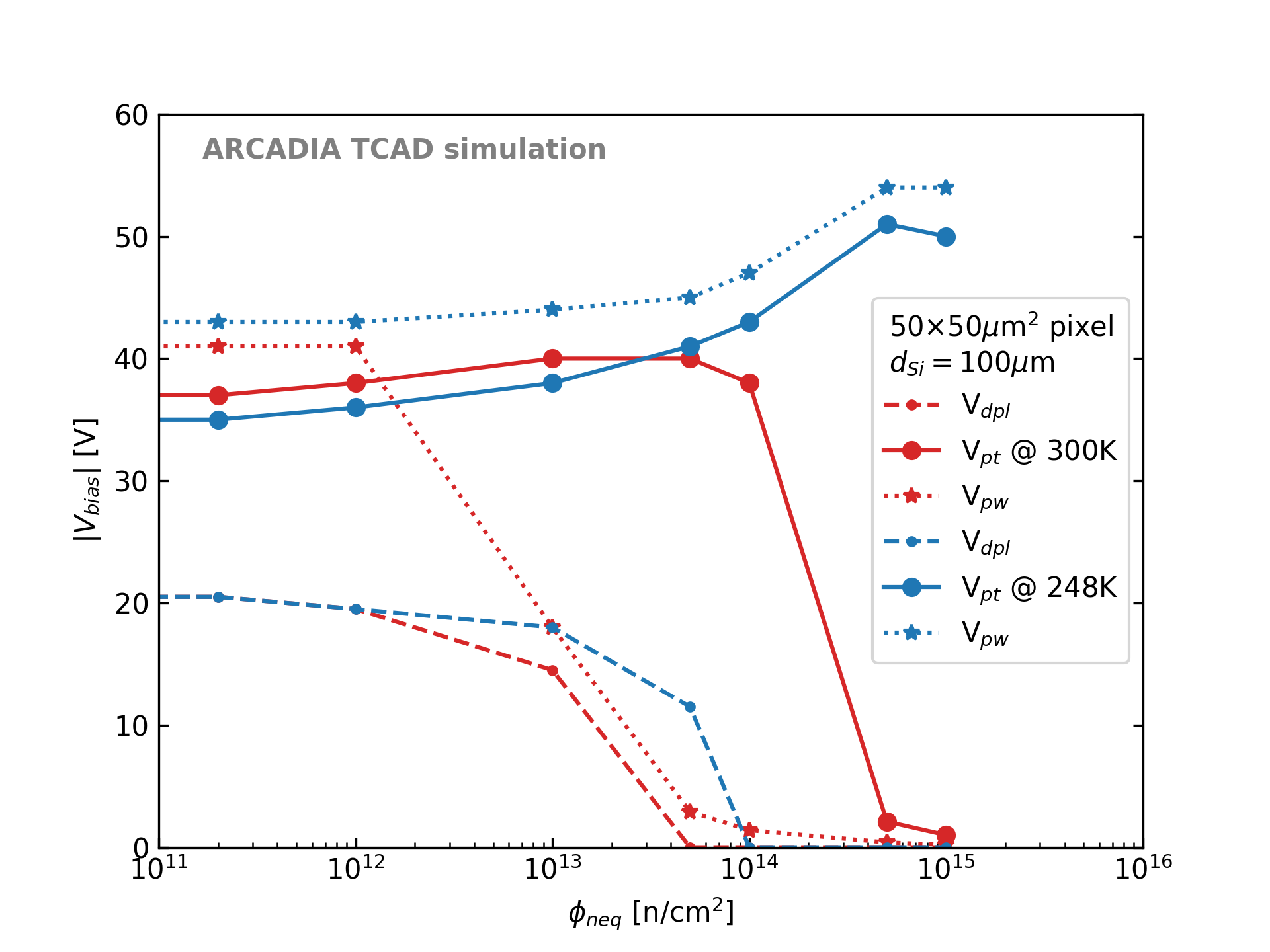}
\caption{Operating voltage of 50$\times$50\,$\mu$m$^2$ pixel in $d_{Si}=100$\,$\mu$m thickness as a function of the fluence. The blue curves refer to simulations run at a temperature of 248\,K, while the red curves refer to a temperature of 300\,K.}
\label{fig:Vop_fluence}
\end{center}
\end{figure}

\begin{figure}[tbp]
\begin{center}
\includegraphics[width=.6\textwidth]{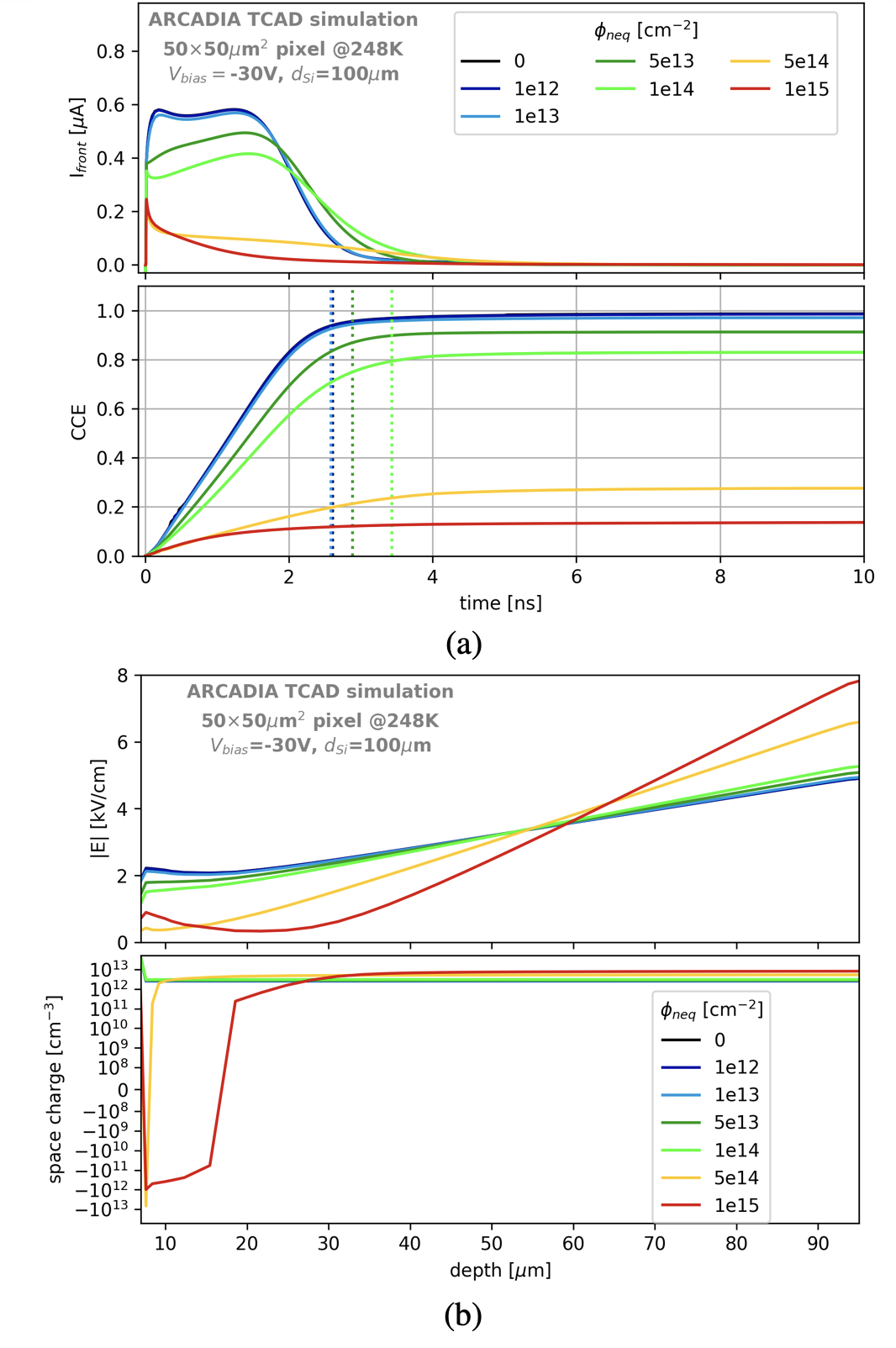}
\caption{(a)~Current signals (top) and charge collection efficiency (bottom) of a 50$\times$50\,$\mu$m$^2$ pixel in $d_{Si}=100$\,$\mu$m thickness, after neutron equivalent fluences of 0 to $1\times10^{15}$\,cm$^{-2}$, simulated at $V_{bias}=-30$\,V. The dotted lines mark the time of 95\% charge collection. (b)~Electric field (top) and space charge (bottom) as a function of the sensor depth for different fluences in the same pixel sensor. }
\label{fig:transient_fluence}
\end{center}
\end{figure}

Evaluating the operating voltages as a function of the fluence (see Fig.~\ref{fig:Vop_fluence}) a drop in the depletion voltage is visible for fluences $>1\times10^{13}$\,cm$^{-2}$. This is consistent with a change of the effective doping concentration within the Si bulk, due to the introduced trap levels. The added acceptor levels first neutralize the n-type doping of the bulk and, with increasing concentration, invert the doping to p-type. 
Consequently, the p-n junction is moved from the back side to the border of the epi layer and the bulk to the front side of the sensor. \\
In conventional Si diodes, the depletion voltage is expected to increase after the inversion of the effective doping of the silicon bulk~\cite{Wunstorf:153817}. However, for these pixel sensors, the depletion voltage is defined as the reverse voltage at which neighboring pixels become isolated by a depletion region, which corresponds to full bulk depletion before irradiation. But as soon as the bulk is type-inverted, the depletion region starts to grow from the front side between the epi layer and the bulk; therefore already with the small applied voltage of $0.8\,V$ at the collection electrode, the electrodes of the pixels become isolated without any reverse bias applied. \\
The type-inversion of the sensors above fluences of $1\times10^{14}$\,cm$^{-2}$ can be confirmed comparing transient simulations of Minimum Ionizing Particles (MIPs) using the \textit{HeavyIonModel} of TCAD~\cite{sentaurus}. This model allows to simulate the charge deposition along a track within the sensor, defining the amount of generated charge as a function of the path length by the linear energy transfer (LET). Complementary \textsc{Geant4} simulations have been used to confirm the average charge deposition of MIPs of $\sim80$\,e/h pairs or $1.28\times10^{-5}$\,pC per $\mu$m.
The charge deposition is instantaneously introduced after 10\,ps, and the resulting transient signal is shown in the top plot of Fig.~\ref{fig:transient_fluence}~(a) for a MIP incident in the pixel center, along a perfectly perpendicular trajectory, for fluences from zero up to $1\times10^{15}$\,cm$^{-2}$. The transients have been simulated at $V_{bias}=-30$\,V, thus well above the depletion voltage over the full fluence range, and the time necessary to reach 95 and 99\,\% charge collection efficiency ($t_{95}$ and $t_{99}$), respectively, are reported in the plot on the bottom. In order to avoid inaccuracies of the simulation that impact the results, these times are defined as:
\begin{equation}
\frac{\int_{t}^{t_{95/99}} I_{front}\cdot dt}{\int_{t}^{t=200ns} I_{front}\cdot dt}=0.95/0.99
\end{equation}
At low fluences $<5\times10^{13}$\,cm$^{-2}$ the pulse shapes present a fast component due to the quickly collected electrons near the collection electrode, and a second peak due to the strong electric field at the pn junction on the back side of the sensor, which results in a higher drift speed of the charges generated close to the back side compares to charged originating at the pixel center. The electric field at $V_{bias}=-30$\,V is shown in the top plot of Fig.~\ref{fig:transient_fluence}~(b) at different fluences. After moderate irradiation of up to $1\times10^{14}$\,cm$^{-2}$, the electric field at the front becomes weaker, and the electrons get trapped by the increasing concentration of acceptor levels. For the highest fluences, a critical amount of the acceptors gets ionized and thus contributes to the space charge, as shown in the bottom plot in  Fig.~\ref{fig:transient_fluence}~(b). The space charge becomes negative below the electrode, where the electron density is highest, which results in a double peak in the electric field~\cite{CHIOCHIA200651}. Additionally, the total collected charge significantly decreases at these extreme fluences due to the high concentration of donor as well as acceptor traps. The hole contribution to the signal seems to vanish completely after a fluence of $\phi_{neq}=1\times10^{15}$\,cm$^{-2}$ and the total collected charge lies below 20\,\%. \\
The occupation of traps is dependent on the amount of available free charge carriers, as well as on the leakage current, the applied bias voltage, and the capture cross-sections. This results e.g. in a stronger double junction at higher temperatures due to easier occupation of the traps and higher leakage currents. Since the employed trap model has been fit to data at 248\,K~\cite{Passeri:2705944}, we report here only the results at 248\,K. \\
It should be mentioned that the amount and the type of radiation induced bulk defects depends on the particle type as well as on the particle energy of the irradiation~\cite{KUHNKE2002144}. In addition, the original impurities within the Si, like oxygen, can support or prevent the creation of certain defects~\cite{Neubuser:2013yla, JUNKES2013113}.

\section{Design optimizations}
\label{subsec:opt}
The pixel design optimization targets a small pixel capacitance to ensure low electronics noise, thus maximum signal to noise ratio, and high charge collection efficiencies within smallest possible time windows in order to decrease dead-times for high particle rates. \\ 
To choose only a few geometries for each pixel pitch (10\,$\mu$m, 25\,$\mu$m, and 50\,$\mu$m) with optimal performance, a large range of simulations with different electrode and pwell sizes has been performed. The results of this simulation campaign are summarized in the following in three different categories, focussing 1) on the pixel capacitance (Sec.~\ref{subsec:minC}), 2) on the bias voltage operating range (Sec.~\ref{subsec:Vop}), and 3) on the charge collection (Sec.\ref{subsec:CC}).\\
 
\subsection{Minimizing the pixel capacitance}
\label{subsec:minC}
\begin{figure}[htbp]
\begin{center}
	\includegraphics[width=\textwidth]{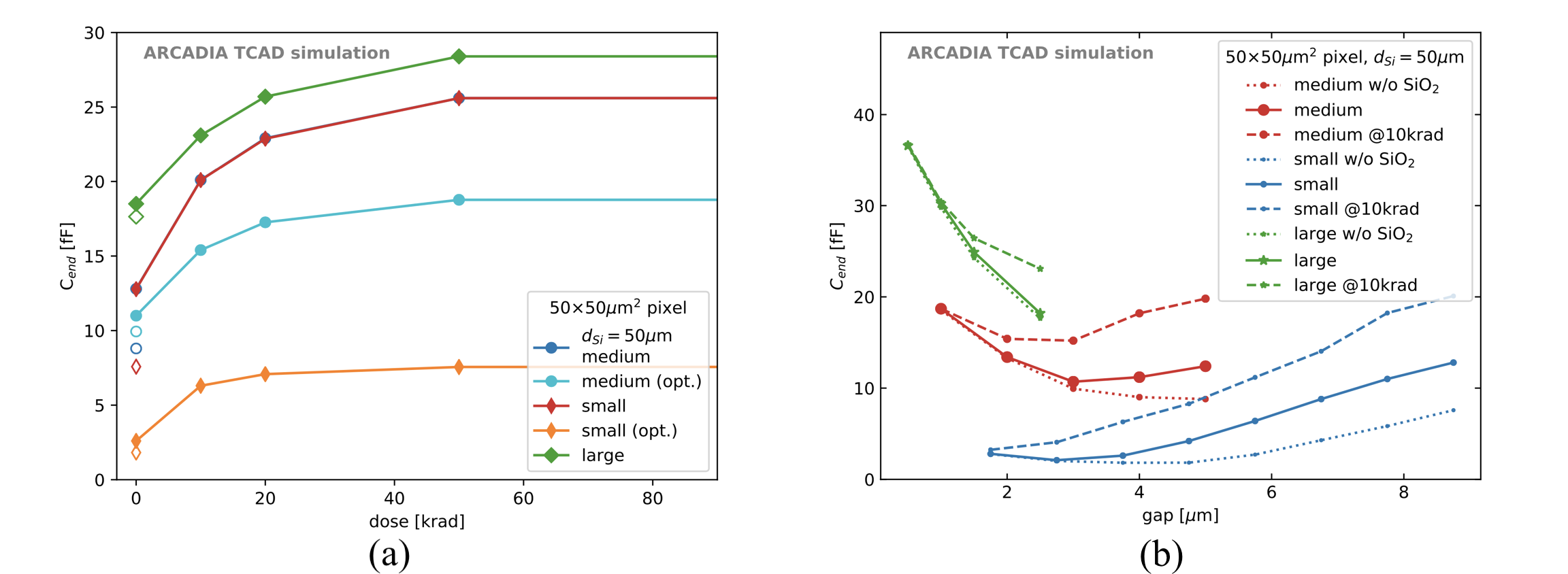}
\caption{(a)~Capacitance of 50$\times$50\,$\mu$m$^2$ pixels in $d_{Si}=50$\,$\mu$m thickness, as a function of the total ionising dose for five different pixel sensor layouts. The dots, diamonds, and triangles correspond to the results obtained for a medium, small, and large collection electrode size, respectively. The light blue dots and orange diamonds show the results after the optimisation of the gap size between the pwell and the electrode for minimal capacitance. The open markers correspond to the simulation results without the SiO$_{2}$ layer. (b)~The capacitance of 50$\times$50\,$\mu$m$^2$ pixels in $d_{Si}=50$\,$\mu$m thickness, with small, medium, and large collection electrode size as a function of the gap size between the pwell and the electrode. }
\label{fig:optC}
\end{center}
\end{figure}

The pixel capacitance can be lowered by decreasing the size of the collection electrode. However, due to the impact of the positive oxide charges within the gap between the electrode and the pwell, the gap size plays a major role and both quantities have to be analyzed in parallel. In the following, bulk damage is not considered in the comparisons between pixel designs, due to a negligible impact on the pixels' capacitance. \\
An example of the importance of the gap size is visible in Fig.~\ref{fig:optC}~(a). The capacitance is shown as a function of the dose for small, medium, and large electrode sizes with default and optimized gap sizes. Here the dose range has been limited to (0 - 100)\,krad, which covers the expected doses reached after a few years of operation within a medical scanner used for proton CT. Comparing the small electrode with the default gap size shows, without the recognition of the SiO$_{2}$ layer and without positive oxide charges, a smaller capacitance than the medium electrode with optimized gap size. However, as soon as oxide charges are introduced, the trend turns around and the capacitance of the medium electrode stays $\sim$30\,\% lower than the capacitance of the non-optimized small electrode. While Fig.~\ref{fig:optC}~(a) only shows the results for the 50\,$\mu$m thick sensor, the same effect and very comparable absolute values have been found for thicker substrates. \\
The capacitance for three different electrode sizes is shown as a function of the gap size for 50$\times$50\,$\mu$m$^2$ pixels in Fig.~\ref{fig:optC}~(b). Due to boundary conditions, the pwell has a minimum size to facilitate the integration of transistors for the electronics; hence the gap size range is limited and most restricted for the largest electrode. For the 50$\times$50\,$\mu$m$^2$ pixel, we find a minimum capacitance for all electrode sizes at $\sim2.5$\,$\mu$m gap size, for simulations that include the SiO$_{2}$ layer. In case of the large electrode size, it can be clearly seen that the capacitance decreases with increasing gap size, which corresponds to a smaller pwell size.\\

\subsection{Operating bias voltage}
\label{subsec:res_Vop}
\begin{figure}[htbp]
\begin{center}
	\includegraphics[width=\textwidth]{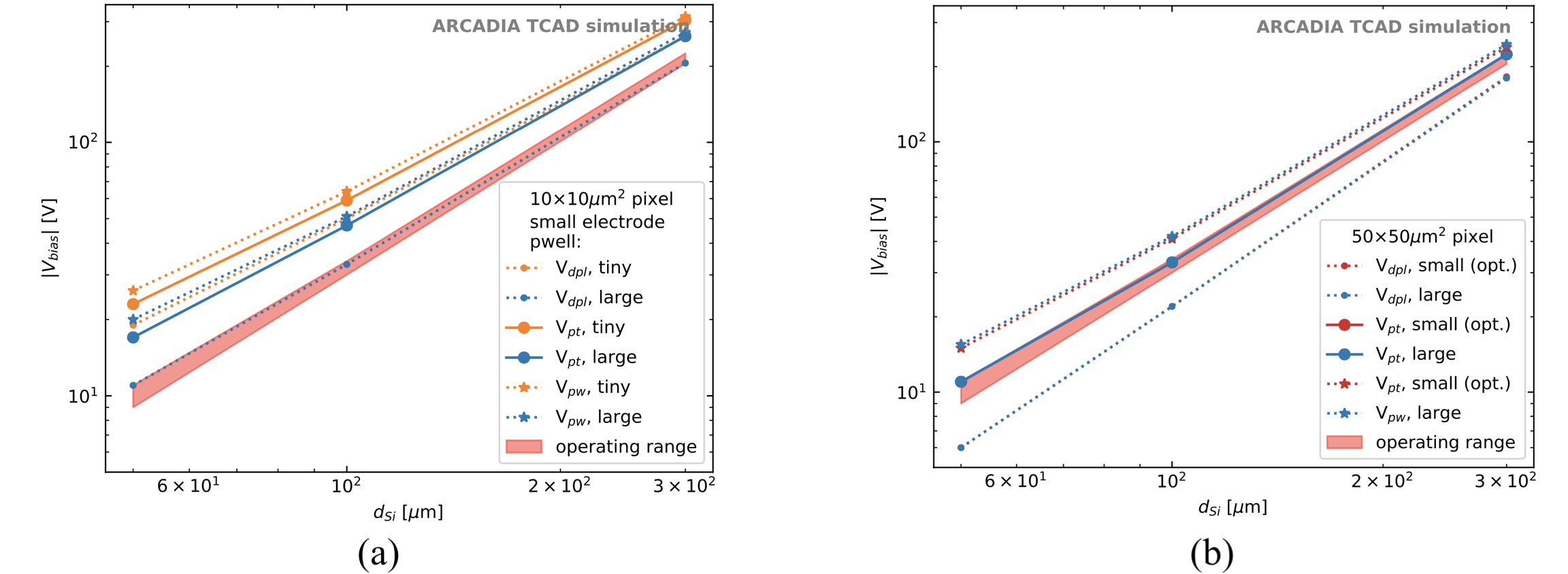}
\caption{Bias voltage at depletion $V_{dpl}$ (small dots), at the onset of the punch-through currents $V_{pt}$ (large dots) and at the point of maximum power consumption $V_{pw}$ (stars), for 10$\times$10\,$\mu$m$^2$ pixels~(a) and 50$\times$50\,$\mu$m$^2$ pixels~(b). The red shaded area corresponds to the $V_{pt}\pm10\%$ of the reference 25$\times$25\,$\mu$m$^2$ pixel.}
\label{fig:opV}
\end{center}
\end{figure}

The pixel designs for the different pitches impact these parameters due to changes in the electrostatic potential. Since the pixel design for the 25\,$\mu$m pitch of the ARCADIA main demonstrator chip is fixed, its operating parameters are used as reference to allow the operation of different pixel flavors on the test-structure. \\
Figure~\ref{fig:opV} shows the bias voltages ($V_{dpl}$, $V_{pt}$, and $V_{pw}$) for the three different thicknesses of the 10$\times$10\,$\mu$m$^2$ pixels (a), and 50$\times$50\,$\mu$m$^2$ pixels (b). The red shaded area marks the $V_{pt}\pm10\%$ of the reference 25$\times$25\,$\mu$m$^2$ pixels. While the reference voltage lies nicely within the limits of all 50$\times$50\,$\mu$m$^2$ pixel designs, the minimum bias voltage of the 10$\times$10\,$\mu$m$^2$ pixels varies strongly for the different designs. Only one design, with a large pwell, can be considered operational (depleted) at the reference voltage. \\
The large difference in the operating bias voltage for the 10$\times$10\,$\mu$m$^2$ pixels has been an important observation and led to the conclusion that the test-structures with these pitches will be separated from the others in the final layout. This will ensure that a sufficient bias voltage can be applied from the front and back side of the sensors for thicknesses of 50\,$\mu$m, and 100/300\,$\mu$m, respectively. \\

\subsection{Uniform response}
\label{subsec:CC}
\begin{figure}[htbp]
\begin{center}
	\includegraphics[width=.59\textwidth]{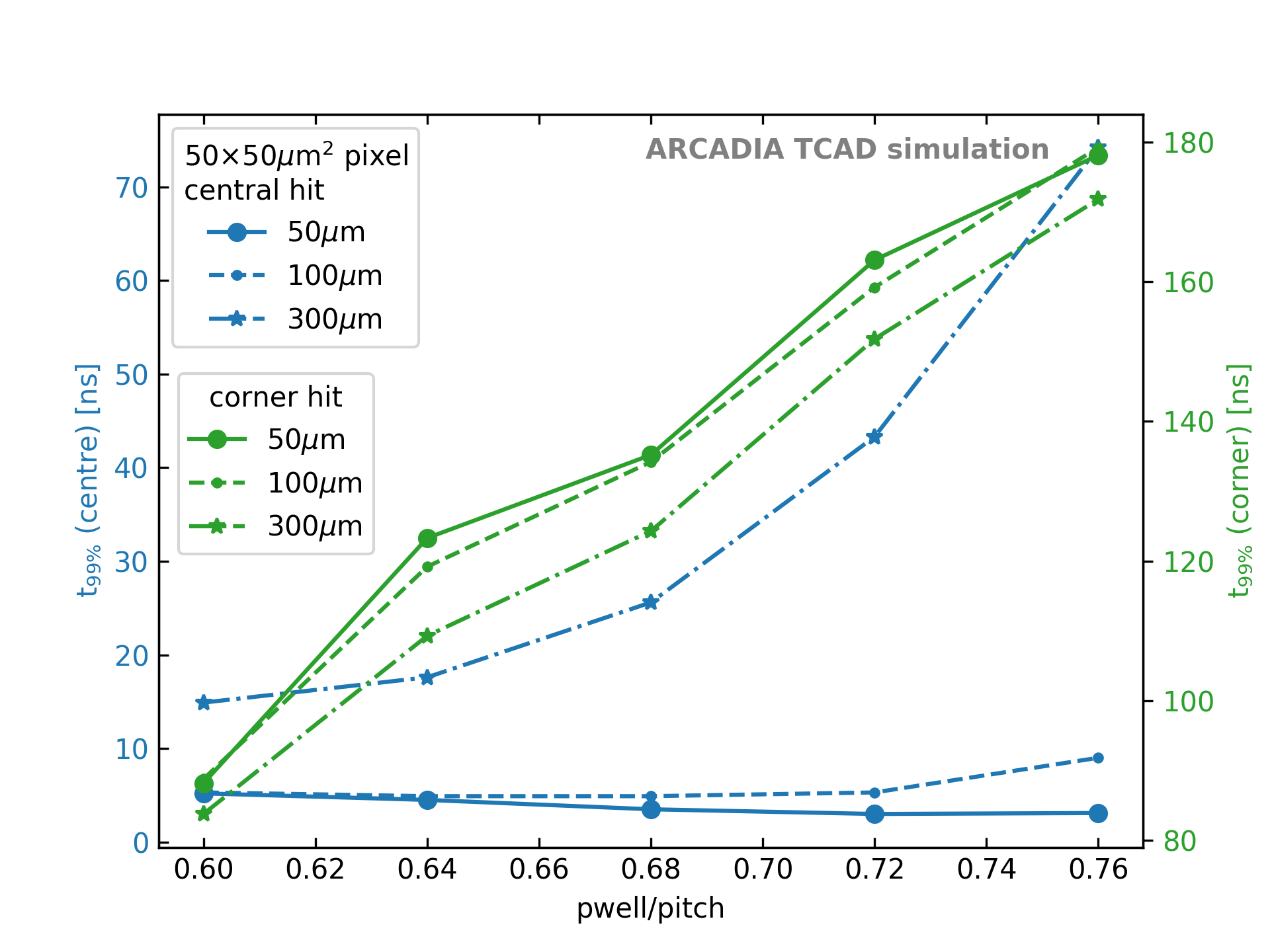}
\caption{Time required for 99\,\% charge collection in the center (blue) and corner (green) of 50$\times$50\,$\mu$m$^2$ pixels with medium sized nwell, as a function of the pwell over pitch ratio. The results are shown for three different sensors thicknesses.}
\label{fig:time_pwell}
\end{center}
\end{figure}

Transient simulations of MIP like tracks have been used to evaluate the charge collection and response across the pixels. To evaluate the signals usually the best and worst case scenarios are studied, which correspond to a traversing particle at the center and the corner of the pixel. For the comparison of different pixel designs, the time to collect 95\,\% and 99\,\% of the generated charges has been used. While charge losses due to radiation damage in the silicon crystal are not expected to vary between the pixel designs but rather with the substrate thickness, the study of response uniformity is neglecting bulk damage. The impact of surface damage on the signal formation has been tested in transient simulations, and has been found to be negligible. \\
A general observation of the necessary time to collect 99\,\% of all charges in the pixel corner is the strong dependence on the pwell width (Fig.~\ref{fig:time_pwell}). Additionally, a large collection electrode size improves the collection of the charges from the corners of the pixels. However, the electric field strength below the collection electrode decreases with an increased collection node, which results in a slightly slower collection in the pixel center. 
Considering a homogeneous distribution of particles across the full area of the chip, motivates an optimization for uniformity rather than peak performance in localized areas. 

\begin{figure}[htbp]
\begin{center}
	\includegraphics[width=.95\textwidth]{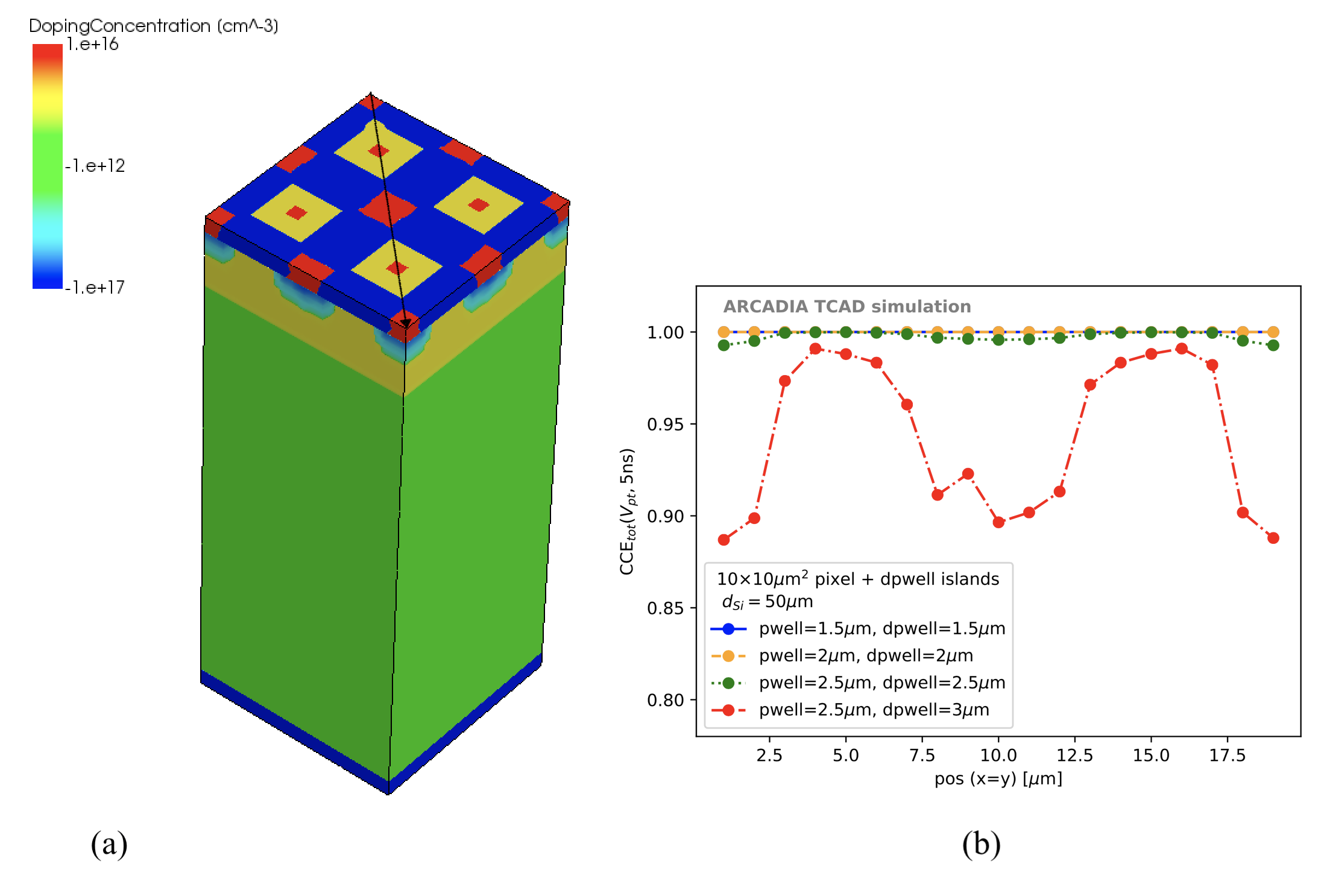}
\caption{(a)~2$\times$2 pixel matrix of 10$\times$10\,$\mu$m$^2$ pixels with dpwell islands in the pixel corners. The black arrow indicates the direction of the position scan. (b)~Charge collection efficiency after 5\,ns for 10$\times$10\,$\mu$m$^2$ pixels with deep pwell islands and $d_{Si}$=50\,$\mu$m, presented as a function of the position along the diagonal indicated in (a). The differently colored curves correspond to different (d)pwell sizes, which values are given in the legend. 
}
\label{fig:CC}
\end{center}
\end{figure}

These considerations lead to a choice of 25$\times$25\,$\mu$m$^2$ and 50$\times$50\,$\mu$m$^2$ pixels, with minimum pwell and maximal electrode size. Due to the small pitch, the 10$\times$10\,$\mu$m$^2$ pixels are much less effected by slow charge collections in the pixels corners. Instead charge sharing is much more pronounced and the larger concern lies in the so-called \textit{channel choking}, which describes the effect of a potential barrier occurring when the sensing window becomes too small in comparison to the pwell size. To overcome this problem, a design that features the electronics in islands of deep pwells within the pixel corners, has been studied and a picture of a $2\times2$ pixel matrix is shown in Fig.~\ref{fig:CC}~(a). With this approach, a large enough (deep) pwell size can be achieved to allow for the integration of the electronics without the risk of a non-functional diode. \\
The resulting charge collection efficiencies for a range of (deep) pwell sizes, given here in half-sizes, are presented in Fig.~\ref{fig:CC}~(b) at $t=5$\,ns for particles incident along the diagonal of the $2\times2$ pixel matrix. The first observation is that for deep pwell size $<3$\,$\mu$m, the charge collection after 5\,ns is always above 99\,\%. The second observation is that even with a deep pwell size of 3\,$\mu$m, the charge collection still reaches 90\,\% after 5\,ns. This feature of fast and uniform responses motivates further studies towards the integration of fast monolithic pixel designs for applications that require fast timing capabilities. 

\section{Conclusions}
\begin{figure}[htbp]
\begin{center}
	\includegraphics[width=.95\textwidth]{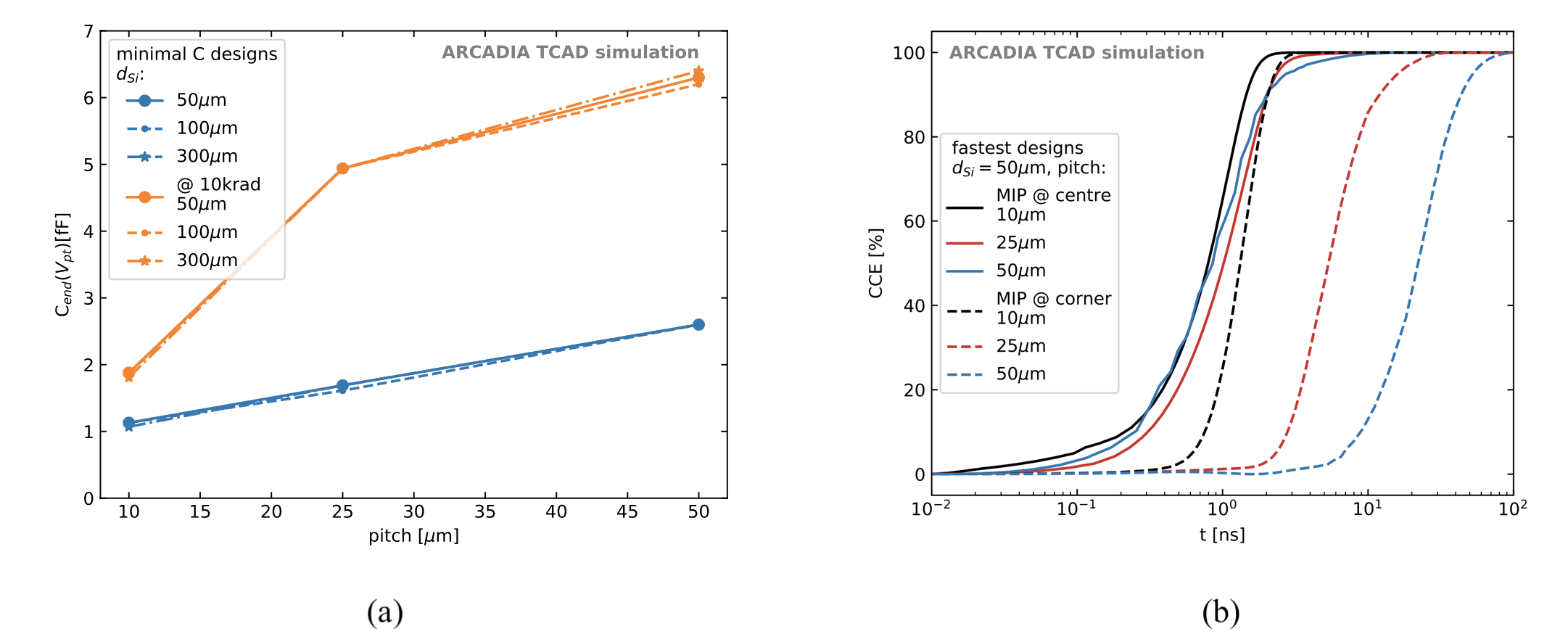}
\caption{(a)~Capacitance at $V_{pt}$ for the optimized designs of pixels with 10, 25, and 50$\mu$m pitch, at dose equal to 0 in blue and after a dose of 10\,krad in orange. (b)~Charge collection efficiencies for MIPs impinging at the center (straight lines) and corner (dashed lines) of the pixels, for the selected design of pixels with 10, 25, and 50\,$\mu$m pitch, before irradiation. }
\label{fig:summary}
\end{center}
\end{figure}

The first engineering run of the ARCADIA main demonstrator is currently in production. The preparations have been accompanied with a large simulation campaign to study possible improvements of the pixel designs in order to decrease the pixel capacitance, and optimize the charge collection. This campaign included first tests of radiation tolerance, using the latest radiation damage models, and has investigated in detail the impact of surface charges on the electrical properties. The capacitance of the selected designs is summarized in Fig.~\ref{fig:summary}~(a), and is shown before and after a dose of 10\,krad is applied. Without surface damage, all pixels feature a single pixel capacitance $<3$\,fF. \\
The charge collection efficiencies for the fastest designs are shown in Fig.~\ref{fig:summary}~(b), for central (straight) and corner (dashed) hits. The times for 99\,\% CC are summarized in Table~\ref{tab:times}. 

\begin{table}[htp]
\begin{center}
\begin{tabular}{c|c|c|c|c|c|c|}
	& \multicolumn{3}{c|}{$t_{99}$(center) [ns]} & \multicolumn{3}{c|}{$t_{99}$(corner) [ns]} \\
pitch	[$\mu$m]& 50\,$\mu$m & 100\,$\mu$m & 300\,$\mu$m & 50\,$\mu$m 	& 100\,$\mu$m	& 300\,$\mu$m \\
\hline
10	& 	2.1		& 3.2			& 7.0 		& 2.8 		& 3.6			& 6.9 \\
25	&	3.6		& 8.0			& 16.5		& 27.9		& 25.8		& 21.3 \\
50	&	7.7		& 6.3			& 14.9		& 73.2 		& 74.3 		& 82.6 \\
\end{tabular}
\end{center}
\caption{Times for 99\,\% charge collection for MIP-like charge depositions at the center and the corner for pixels with three different pitches, given for three substrate thicknesses (50/100/300)\,$\mu$m. }
\label{tab:times}
\end{table}%

These studies led to the choice of a few different designs of pixels with 10/25/50\,$\mu$m pitch that will be realized in a number of test-structures, such as e.g. matrices of $1\times1$ and $2\times2$\,mm$^{2}$. These structures will allow a validation of the physics models used in the simulation and will be characterized using, besides basic IV and CV measurements, radioactive sources and a laser for e/h pair generation.




\section*{Acknowledgments}
The research activity presented in this article has been carried out in the framework of the ARCADIA experiment funded by the Istituto Nazionale di Fisica Nucleare (INFN), CSN5. 

\bibliographystyle{elsarticle-num} 
\bibliography{bib}

\end{document}